\documentclass[twocolumn]{revtex4-1}          

\def\la{\mathrel{\mathpalette\fun <}}

\def\fun#1#2{\lower3.6pt\vbox{\baselineskip0pt\lineskip.9pt
  \ialign{$\mathsurround=0pt#1\hfil##\hfil$\crcr#2\crcr\sim\crcr}}}

\def\beq{\begin{equation}}
\def\eeq{\end{equation}}
\def\beqar{\begin{eqnarray}}
\def\eeqar{\end{eqnarray}}

\def\iso#1#2{\mbox{${}^{#2}{\rm #1}$}}

\def\li#1{\iso{Li}{#1}}
\def\c1#1{\iso{C}{1#1}}
\def\n1#1{\iso{N}{1#1}}
\def\o1#1{\iso{O}{1#1}}

\begin{document}

\title{Galactic Fly-Bys: New Source of Lithium Production}

\author{Tijana Prodanovi\'{c}}
\email{prodanvc@df.uns.ac.rs}
\affiliation{Department of Physics, Faculty of Sciences, University of Novi Sad \\
Trg Dositeja Obradovi\'{c}a 4, 21000 Novi Sad, Serbia}

\author{Tamara Bogdanovi\'{c}}
\altaffiliation{Center for Relativistic Astrophysics, Georgia Tech}
\affiliation{School of Physics, Georgia Institute of Technology\\
837 State Street, Atlanta, GA 30332, U.S.A.}

\author{Dejan Uro\v{s}evi\'{c}}
\affiliation{Department of Astronomy, Faculty of Mathematics, University of Belgrade \\
Studentski Trg 16, 11000 Belgrade, Serbia}

\begin{abstract} 
Observations of low-metallicity halo stars have revealed a
puzzling result: the abundance of \li7 in these stars is at least
three times lower than their predicted primordial abundance. It is
unclear whether the cause of this disagreement is a lack of
understanding of lithium destruction mechanisms in stars or the
non-standard physics behind the Big Bang Nucleosynthesis (BBN). 
Uncertainties related to the destruction of lithium in stars can be
circumvented if lithium abundance is measured in the "pristine" gas of
the low metallicity systems.
The first measurement in one such system, the Small Magellanic Cloud (SMC), was found to be at the level
of the pure expected primordial value, but is on the other hand, just barely consistent with the expected galactic abundance
for the system at the SMC metallicity, where important lithium quantity was also produced in interactions of
galactic cosmic rays (GCRs) and presents an addition to the already present primordial abundance.
Due to the importance of the SMC lithium measurement for the resolution of the
lithium problem, we here draw attention to the possibility of another
post-BBN production channel of lithium, which could present an important addition 
to the observed SMC lithium abundance. 
Besides standard galactic cosmic rays, additional post-BBN production
of lithium might come from cosmic rays accelerated in
galaxy-galaxy interactions. This might be important for a system such is the SMC, which
has experienced galaxy harassment in its history.
Within a simplified but illustrative framework we demonstrate that large-scale tidal
shocks from a few galactic fly-bys can possibly produce lithium in amounts
comparable to those expected from the interactions of galactic
cosmic-rays produced in supernovae over the entire history of a
system. In case of the SMC, we find that only two such fly-bys could possibly
account for as much lithium as the standard, GCR production
channel. However, adding any a new mechanism for
post-BBN production of lithium, like the one proposed here, would 
contribute to the observed SMC lithium abundance, causing this
measurement to be more in tension with the primordial abundance
predicted by the standard BBN.
\end{abstract}

\maketitle

\section{Introduction}

One of the key tests of the hot big bang model are predictions of the
primordial abundances of light elements, made in the Big Bang
Nucleosynthesis (BBN).  The discovery of the lithium abundance plateau (a
uniform, metallicity independent, lithium abundance) measured in 
low-metallicity halo stars \citep{spite} indicated that primordial
abundance had been observed. However, in the past decade, it became evident
that primordial lithium abundance, $(\li7/\rm H)_p = 5.24
\times 10^{-10}$ \citep{cfo08}, predicted in the standard Big Bang
Nucleosynthesis framework and calibrated by the cosmic microwave
background observations \citep{wmap09}, is a factor of $2-4$ higher
than the observed plateau value, $(\li7/\rm H)_{plateau} = 1.23 \times
10^{-10}$ \citep{ryan}. This is commonly referred to as the lithium problem.

Recently, more extensive observational surveys,
higher resolution spectra, and improved stellar modeling have revealed
more complexity in the appearance of the ``Spite plateau".  They
indicate a greater dispersion in lithium abundance below
metallicity $\left[ \rm Fe/H \right] \la -3$, where lithium depletion
levels show significant variations from star to star
\citep{sbordone2012,aoki2009,melendez2010}. The notion of a plateau 
has consequently been replaced by an upper envelope of lithium
abundance at the level of $(\li7/\rm H)_{plateau}$ for stars with
$\left[ \rm Fe/H \right] \la -1.5$. Very few outliers have been
reported to lie in the ``forbidden zone" above this envelope
\citep{spite2012,iocco2012a,fields2011}.

In addition to BBN, \li7 is also produced in cosmic-ray interactions
\citep{reeves} and by the neutrino process \citep{ww}. 
In fact, most of the light isotope \li6 observed in the present
epoch ($(\li6/\rm H) \sim 10^{-10}$) was made by interactions of cosmic
rays with the interstellar medium (ISM)\citep{reeves}. Smaller amounts
of \li6 ($(\li6/\rm H)_p \sim 10^{-14}$) were also created in the BBN
\citep{thomas1993,vangioni1999} and possibly, as recently pointed out, 
in accretion processes \citep{iocco2012b}.
Furthermore, as supernova remnants are thought to be the dominant
galactic source of cosmic rays (GCRs), \li6 abundance is expected to
increase with metallicity. This prediction has been challenged by
reports of a tentative \li6 plateau in low-metallicity halo stars
\citep{asplund}. Since then, several of the originally reported 
``plateau values'' of \li6 abundance have been revised after 
improved 3D non-LTE modeling
\citep{lind2012,asplund2010,prantzos2012,steffen2012,cayrel2007}. 
However, at least two anomalously high \li6 measurements remain, and, if
confirmed, their explanation would require an additional, non-standard
source of \li6, and consequently \li7.

One possible solution to this puzzle may be in the form of the
non-standard BBN \citep{jedamzik, jedamzik2009,iocco2009,cyburt2009}.
Alternatively, one could appeal to early cosmic-ray populations
different from standard GCRs \citep[for e.g.,][]{si,pf07}.  A
difficulty encountered by all models is that they fail to produce
significant amounts of \li6 without violating metallicity or
energy constraints, and overproducing other light elements \citep{prantzos}. 
In order to establish levels
of \li6 and \li7 which are uncomplicated by the details of processing in stellar
atmospheres, it has been proposed that their abundances be measured
in the pristine and unprocessed gas of low metallicity systems
\citep{pf04,vidal-madjar1987,steigman1996}. The first such observation of
gas phase lithium beyond our galaxy has been carried out recently
in the Small Magellanic Cloud (SMC), and is an important step towards finding the cause
of the observed discrepancy between expected primordial abundance and that
measured in low-metallicity halo stars. 
It revealed a value of
lithium abundance, $(\li7/\rm H)_{SMC} = 4.8 \times 10^{-10}$, which
is consistent with the primordial value \citep{howk}. On the other hand, in the systems at 1/5 of solar metallicity,
such is the case with the SMC, some, non-negligible post-BBN production of lithium is also expected, due to
interactions of GCRs with the gas in the interstellar medium. This would add to the already present primordial abundance, and
be included in the observed value. Hence, at 1/5 solar metallicity, total lithium abundance in the gas phase of this system should be 
higher than the primordial. In the case of the SMC, the observed abundance is just marginally consistent with expected abundance
for the system at given metallicity. Therefore, while a new lithium source (different from the
GCRs) may be needed to explain the \li6 excess in some systems, SMC measurement
leaves little room for any non-standard post-BBN source which would yield 
significant amount of \li7. Consequently, if any additional significant source of lithium is present, 
the current SMC measurement would then become inconsistent with the expected abundance (BBN + post-BBN production), just
is the case with lithium measured in atmospheres of low-metallicity halo stars.

In this work, we point out that tidal cosmic rays (TCRs) could be a
significant source of lithium in systems that have undergone strong
tidal interactions with their neighbors.  
If present, this could be a source of lithium that has not been previously taken into account
but might result in important consequences.
Close halo fly-bys play an
important role in the evolution of the earliest dark matter halos and
their galaxies, and can still influence galaxy evolution in the
present epoch \citep[for e.g.,][]{sinha}.  Galactic mergers and close
fly-bys are known to give rise to large-scale shocks in the gas of
interacting galaxies \citep{toomre, barnes92, barnes96, mihos96,
moore, cox06}. These shocks are favorable locations for
acceleration of cosmic rays, which in turn can produce
lithium. However, shocks triggered by galaxy interactions are not directly
accompanied by fresh metal yields and could in principle circumvent
the problem of overproduction of metals faced in other models. They
can nevertheless be indirectly accompanied by fresh metal yields, as
galaxy interactions are known to enhance star formation
\citep{mihos94, hernquist, cox08, bekki}. If so, tidal
cosmic-ray populations may be accompanied by some increase in
metallicity, but the correlation would be weaker than in supernovae,
which eject fresh metals and accelerate particles at the same time. 

At high redshift, where destructive interactions of comparable mass
galaxies were more common, the TCRs may have competed with the GCRs
accelerated by the first generation of massive stars in the production of
light elements. At low redshift, TCR nucleosynthesis could be
important for low metallicity systems, which continue to experience
tidal disruptions by their neighbors, such as the SMC \citep[see
eg.][]{diaz,yoshizawa,connors}.  In these systems, at a given
metallicity, one would thus expect to find a significantly higher \li6
abundance and consequently, a lower \li7-to-\li6 ratio relative to
that predicted by standard galactic chemical evolution models.  If
the Milky Way (MW) has not suffered a major tidal encounter with its
neighbors at high redshift, TCRs may not have contributed much to the
lithium measured in halo stars. We propose that this effect may be
important for the SMC, which is actively interacting with the MW and
the Large Magellanic Cloud (LMC), and that lithium abundance
measurements in these galaxies should reflect their different
evolutionary paths.

Using a simple analysis, we show that the energy imparted by galactic
tidal encounters is sufficient to produce significant lithium
abundance. We also find that only a few galactic fly-bys can yield large enough TCR fluxes which could result in
lithium amounts comparable to those produced by the GCRs over the entire
history of a galaxy. Finally, in case of the SMC, we show that its gas
phase lithium abundance could have been significantly enriched in
tidal encounters with its immediate neighbors, the Milky Way and the
Large Magellanic Cloud.
Thus, the two main objectives of this work are: 1. to point out
to a new cosmic-ray population which may arise in interacting systems and
have important consequences for nucleosynthesis and expected gamma-ray and radio emissions, 2.
to draw attention to the fact that though extremely important, the SMC lithium
gas-phase measurement currently does not provide a definitive answer about the lithium problem,
and that any additional, significant post-BBN production of lithium can tip the scale, thus having
important consequences for the further analysis of this problem.
 
\section{Energetics}

In order for galactic interactions to be a viable source of energy for
production of Li in metal poor environments, they have to satisfy two
important criteria: (1) the energy released in large scale tidal
shocks should account for the energy necessary to produce the level of
Li measured in these systems, and (2) tidal shocks must be capable of
accelerating a population of cosmic rays responsible for Li production.  In
this section we place an upper limit on the energy available for
nucleosynthesis, by estimating the kinetic energy of the encounter for
fiducial parameters representative of a minor encounter of a primary
galaxy with its less massive satellite. The available energy can be
estimated as
\beqar
\label{eq_kinetic}
E_{\rm kin} &=& \frac{q\,G\,M_1^2}{d}   \\
\nonumber
& \approx &  4\times 10^{57} {\rm erg}\, 
\left(\frac{q}{10^{-3}}\right)  
\left(\frac{M_1}{10^{12}M_\odot}\right)^2 
\left(\frac{d}{50\,{\rm kpc}}\right)^{-1}
\eeqar
where $G$ is the gravitational constant, $q = M_2/M_1 < 1$ is the mass
ratio of the satellite to primary galaxy, and $d$ is their
separation. Note that the expression for kinetic energy is evaluated
for a satellite galaxy plunging toward the primary on a nearly radial,
marginally gravitationally bound orbit. As indicated by simulations of
galactic mergers, this type of encounter is typically more damaging for
the satellite galaxy which is tidally stripped of its mass as it falls
into the larger galaxy \citep{callegari09, callegari11}. Because of
its shallower potential well, the gas in the satellite galaxy which is
not lost to tidal stripping can be strongly shocked, even though the
satellite may inflict little damage to its host. The shock is
expected to be more severe for plunging satellites, and, as in this
case, strong perturbation to their potentials occurs rapidly, on a
dynamical time scale. On the other hand, slowly inspiralling
satellites experience changes in their potential over many orbits,
during which the gas and stars gradually adjust to a new
quasi-equilibrium.
 
We estimate the strength of the shocks that arise during the minor tidal
interaction described above by calculating the Mach number of the
interaction for assumed properties of the ISM in the satellite galaxy as
\beqar
\label{eq_Mach}
{\mathcal{M}} &=& \frac{V_{\rm sat}}{c_{\rm s}}  \\
\nonumber
& \approx & 
460\,\mu^{1/2} \left(\frac{M_1}{10^{12}M_\odot}\right)^{1/2} 
\left(\frac{d}{50\,{\rm kpc}}\right)^{-1/2} \left(\frac{T}{100K}\right)^{-1/2}
\eeqar
\noindent where $V_{\rm sat}$ is the infall velocity of the satellite,
$c_{\rm s}$ is the average speed of sound of the ISM gas in the
satellite galaxy, $\mu$ is its mean atomic weight, and $T$ is the
mass weighted average temperature. Note that $T=100K$ corresponds to
cold neutral medium, composed mostly of hydrogen with typical
densities of $20-50\,{\rm cm^{-3}}$. In reality however, the ISM gas
is likely to be a mixture of several phases at different temperatures
\citep{mo77} and this value would vary as a function of satellite
properties and redshift. However, even an order of magnitude increase
in the mass weighted average temperature of the ISM of a particular
satellite would still allow strong shocks to develop as a consequence
of its infall. We will use this robust property of tidal shocks to
constrain the spectrum of the produced cosmic rays that can give rise
to Li formation.

We further estimate what fraction of the kinetic energy in a galactic
encounter is converted into the acceleration of energetic
particles. We assume that the composition of cosmic rays reflects the
composition of the ISM, and consequently, that the $\alpha + \alpha$
fusion channel dominates lithium production at low metallicities
\citep{sw}. This assumption is justified for the low metallicity gas
in the Small Magellanic Cloud, which we employ as a case study in this
work. Following \citet{prantzos} we assume that it takes
$\epsilon_6=16$ erg of energy to produce one nucleus of \li6 via
$\alpha + \alpha$ fusion channel (note that different compositions of
the cosmic-ray population imply different energy requirements per
nucleus). The adopted production energy per nucleus was derived within
the standard "leaky box" framework, where cosmic rays accelerated in
supernova remnants (SNRs) are allowed to escape from the Galaxy and
suffer other losses as they propagate through it. This results in an
equilibrium cosmic-ray spectrum which is steeper at the higher energy end,
and shallower at the low-energy end, relative to the initial injection
spectrum produced at the location of the strong supernova shocks.
Given the high Mach number value estimated in equation
(\ref{eq_Mach}), which falls within the wider range of values
characteristic for supernovae shocks, we assume that tidal shocks
with $\mathcal{M}>100$ will have cosmic-ray injection spectrum
similar to the injection spectra from supernovae. Subsequently, the
tidal cosmic-ray population is expected to suffer similar loses during
TCR propagation through the galaxy, resulting in an equilibrium
spectrum similar to that of galactic cosmic rays.  This is the key
assumption (see discussion in Section~\ref{S_discussion}) which will
later allow us to evaluate the efficiency of TCR nucleosynthesis
relative to GCR nucleosynthesis, without making explicit choices
for the (unknown) TCR spectrum.

It is worth noting though that the uncertainty involved in the nature
and evolution of the TCR spectrum is somewhat offset by the fact that
the adopted energy per \li6 nucleus is less sensitive to a specific
particle acceleration mechanism and can be applied to a wide range of
acceleration scenarios \citep{prantzos}. Expressed per gram of ISM
matter, this energy requirement is
\beq \omega_6=\epsilon_6\, y_6
\frac{1}{m_{\rm p}} = 1.5 \times 10^{15} {\rm erg\,gr^{-1}} \left(
\frac{\epsilon_6}{16 \rm erg} \right) \left(
\frac{y_6}{y_{6, \odot}} \right) 
\eeq 
where $m_{\rm p}$ is the proton mass, while the solar abundance of
\li6 is $y_{6, \odot} \equiv (\li6 /{\rm H})_\odot=1.53 \times
10^{-10}$ \citep{ag}. The total energy required to pollute the amount
of gas $M_{gas}$ with lithium abundance $y_6$ is
\beq E_6 =
\omega_6 M_{\rm gas} = 3 \times 10^{57} {\rm erg} \left(
\frac{\epsilon_6}{16 \rm erg} \right)  \left( \frac{y_6}{y_{6,
\odot}} \right) \left( \frac{M_{\rm gas}}{10^9 M_\odot} \right)
\label{eq:energy_li6} 
\eeq
The derived value of energy implicitly depends on the assumed
cosmic-ray spectrum, escape length, and metallicity (through the
choice of energy-per-nucleus); we discuss the importance of these
parameters in Section~\ref{S_discussion}.

\section{Requirements for Significant Lithium Production}

Tidal shocks that arise from close galactic fly-bys can accelerate
charged particles and in such way as to give rise to a new cosmic-ray
population within an interacting galaxy. While standard GCRs are
expected to be produced in SNRs over the entire history of a galaxy,
tidal cosmic rays are injected in the interstellar medium
episodically, and only during sufficiently strong tidal events
($\mathcal{M}> 100$), as indicated by the Mach number of the
encounter. After the point of closest approach in a fly-by, the TCR
flux is likely to rapidly decrease due to energy losses, and 
subsequent nucleosynthesis would stop.  As tidal shocks in galactic
fly-bys can affect much larger ISM volumes than supernovae shocks,
they can in principle compensate for their low ``duty cycle'' by their
high volume feeling fraction. Whether the GCR or TCR driven
nucleosynthesis dominates in a given galaxy depends on the parameters
of the encounter and properties of the interacting galaxies. Modeling
such encounters requires high resolution hydrodynamic simulations to
capture the structure of the tidal shocks, and is beyond the scope of
this paper. Instead, we focus on the question of whether cosmic rays
accelerated in tidal shocks are a plausible and important source of
lithium in galaxies which have experienced close encounters in their
history.

We assume that tidal shocks propagate through the magnetized ISM of
the satellite galaxy, causing perturbations in its magnetic field, and
accelerating charged particles. This is similar to the diffusive shock
acceleration of standard GCRs \citep{bell,bo,drury}, which is a
first order Fermi particle acceleration process and a mechanism
routinely adopted in a variety of astrophysical environments.
In addition to first order, second order Fermi particle
acceleration can arise in the downstream region of tidal shocks, although
its contribution to the dominant diffusive shock acceleration process
is likely to be small and negligible \citep{longair}. 

Given the similarity of the acceleration mechanisms and the strength
of the shocks as given by their Mach numbers, we proceed by assuming a
comparable efficiency of TCRs and GCRs in the production of lithium.
We estimate the volume of the ISM in an interacting galaxy that needs
to be shocked in order to give rise to a TCR flux sufficient to
produce an abundance of lithium equal to that produced by GCRs over
the entire history of the system. Thus, we start by equating the total
number of Li nuclei produced by the TCRs and GCRs, $N_{\rm Li, TCR} =
N_{\rm Li, GCR} $. In both cases, the number of Li nuclei can be
expressed in terms of their production rate per unit volume
$\dot{n}_{\rm Li}$ as $N_{\rm Li} = \int \dot{n}_{\rm Li} V_{\rm sys}
dt$, where $V_{\rm sys}$ is the volume in which the CRs interact with
the ISM in each scenario.  The production rate of lithium however
depends on the number density of the ISM ($n_{\rm ISM}$), the cross
section for lithium production in $\alpha + \alpha
\rightarrow \rm Li$ fusion channel ($\sigma$), and on the cosmic-ray flux
($\Phi_{\rm cr}$) as
\beq
 \dot{n}_{\rm Li} = n_{\rm ISM}\, \sigma \Phi_{\rm cr}
\eeq
where $\Phi_{\rm cr} [{\rm cm^{-2} s^{-1}}]= \int \phi (E) dE = \int
v_{{\rm cr},E} (dn_{{\rm cr},E}/dE) dE \propto \int E^{-\alpha} dE$
with cosmic-ray spectral index $\alpha$.  The energy integrated cosmic-ray
flux can also be written in terms of the mean CR velocity and CR
number density as $\Phi = \langle v_{\rm cr} \rangle n_{\rm cr}$.  The
lithium production rate then becomes $\dot{n}_{\rm Li} = n_{\rm ISM}\,
\sigma \langle v_{\rm cr} \rangle N_{\rm cr} / V_{\rm sys}$.  Assuming
that the cosmic-ray flux does not vary much over the production timescale
$\tau_{\rm cr}$, i.e. that the cosmic-ray flux is in equilibrium, the total
number of lithium nuclei produced can now be written as
\beq
 N_{\rm Li} = n_{\rm ISM}\, \sigma \langle v_{\rm cr} \rangle N_{\rm cr} \tau_{\rm cr}.
\eeq
where $N_{\rm cr}$ is the total number of cosmic rays accelerated by a
given process over the entire timescale.  Assuming the same spectral index of
both cosmic-ray populations, mean cosmic-ray velocities $\langle
v_{\rm cr} \rangle= \int v_{{\rm cr},E} (dn_{{\rm cr},E}/dE) dE / \int
(dn_{{\rm cr},E}/dE) dE $ will be equal. It then follows that
\beq
N_{\rm TCR}= N_{\rm GCR,tot}\,\frac{\tau_{\rm GCR}}{\tau_{\rm TCR}} 
\eeq
The two cosmic-ray populations are not actively producing lithium over the
same time-scales. GCRs are producing lithium continuously over the
life time of a galaxy, and we take this timescale ($\tau_{\rm GCR}$)
to be comparable to the age of the Universe, $\tau_{\rm
GCR}=10^{10} \rm yr$ . TCRs, on the other hand,
are accelerated only during close galactic fly-bys, while tidal shocks propagating through the satellite galaxy remain strong.
Their
duty-cycle time scale is comparable to the dynamical time scale for the
interaction of the two galaxies for which we adopt a value $\tau_{\rm TCR} = 10^9$ yr 
(see Section~\ref{S_discussion} for discussion).
It follows that
\beq
N_{\rm TCR}= 10 N_{\rm GCR}\,N_{\rm SN} \left( \frac{10^9 {\rm yr}}{\tau_{\rm TCR}} \right)
\label{eq:fluence}
\eeq
  where $N_{\rm GCR}$ is
the number of cosmic rays accelerated in one SNR and $N_{\rm SN}$ is
the number of supernovae that have occurred up to some epoch, defined by a
given metallicity threshold.  We express the number of cosmic rays
(either TCRs or GCRs) accelerated per fly-by, or in a single SNR, in
terms of the dimensionless injection parameter, $\eta = N_{\rm
acc}/N_{\rm s}$ as defined in \citep{berezhko}, which represents the
number of accelerated particles relative to the number of particles
swept up by the shock. In case of GCRs $\eta_{\rm GCR}=N_{\rm
GCR}/N_{\rm SN,s}$ where $N_{\rm SN,s}$ is the number of particles
swept up by a single supernova shock. In case of TCRs, $\eta_{\rm
TCR}=N_{\rm TCR}/N_{\rm T,s}$, where $N_{\rm T,s}$ is the number of
particles swept up by a tidal shock. Taking these into account we
rewrite equation (\ref{eq:fluence}) as
\beqar
N_{\rm T,s} &=& 10 N_{\rm SN}\,N_{\rm SN,s} 
\left( \frac{\eta_{\rm GCR}}{\eta_{\rm TCR}} \right) 
\left( \frac{10^9 {\rm yr}}{\tau_{\rm TCR}} \right)
\label{eq:sweaptnumber}
\eeqar
While our result does not explicitly depend on the adopted value of
the injection parameter $\eta$, which encodes the acceleration
efficiency, it does depend on the relative efficiency of particle injection
in tidal shocks relative to supernovae shocks.  By adopting
$\eta_{\rm TCR} \sim \eta_{\rm GCR}$ in this estimate, we are making an
implicit assumption that tidal shocks are as strong as supernovae
shocks. In reality, tidal shocks are significantly weaker than the
strong shocks in young SNRs where the velocity of the blast wave can
be as high as $2\times 10^4\,{\rm km\,s^{-1}}$. The velocity of a tidal wave is, however, similar in
strength (as quantified by the Mach number) to shocks driven by the
moderately evolved SNRs sweeping the ISM with velocities $\lesssim
10^3\,{\rm km\,s^{-1}}$. Since weaker shocks are characterized by
slightly higher $\eta$ values \citep{malkov}, our assumption about the
comparable strength of the two types of shocks is conservative. 

The number of particles swept by one supernova can then be estimated
as
\beqar
\nonumber
N_{\rm SN,s} &=& n_{\rm ISM}\,V_{\rm SNR}  \\
&=& 1.2 \times 10^{59} \left( \frac{n_{\rm ISM}}{1 {\rm cm^{-3}}} \right) 
\left( \frac{R_{\rm SNR}}{10 {\rm pc}} \right)^3
\label{equation8}
\eeqar
normalized to fiducial values of the ISM number density $n_{\rm ISM}=1 {\rm
cm^{-3}}$ and the corresponding maximal SNR radius within which
particles are efficiently accelerated. We note, however, that
depending on the energy of the explosion and on the ISM density, the maximal SNR radius
for which the associate shock wave is still capable of accelerating particles to cosmic ray energies,
can be taken to be up to 25pc \citep{berezhko}. 

We now estimate the number of supernova events that occurred by a
certain epoch as determined by the threshold metallicity that these
SNe contributed to the interacting galaxy.  Adopting the solar abundance
of iron $y_{\rm Fe_\odot} \equiv (n_{\rm Fe}/n_{\rm H})_\odot=3 \times
10^{-5}$ \citep{ag} and mass fraction $X_{\rm Fe_\odot} \equiv
(\rho_{\rm Fe}/\rho_{\rm gas})_\odot = 1.25 \times 10^{-3}$, the total
iron mass of such a system is
\beqar
\nonumber
M_{\rm Fe} &=& X_{\rm Fe_\odot} M_{\rm gas} 
\\ &=& 1.25 \times 10^6 M_\odot 
\left( \frac{y_{\rm Fe}}{y_{{\rm Fe}, \odot}} \right) 
\left( \frac{M_{\rm gas}}{10^9 M_\odot} \right) 
\eeqar

We calculate the number of SN events that give rise to the solar
metallicity by adopting a mean iron yield per supernova 
 $M_{\rm Fe, SN} = 0.2 M_\odot$ \citep{pagel}.
\beqar
\nonumber
N_{\rm SN} &=& M_{\rm Fe}/M_{\rm Fe, SN} \\
&=& 6.25 \times 10^6 \left( \frac{0.2 M_\odot}{M_{\rm Fe, SN}} \right) 
\left( \frac{y_{\rm Fe}}{y_{\rm Fe, \odot}} \right) 
\left( \frac{M_{\rm gas}}{10^9 M_\odot} \right) 
\label{eq:nsnumber}
\eeqar

Using equations~(\ref{eq:sweaptnumber}), (\ref{equation8}), and
(\ref{eq:nsnumber}) we write the number of particles swept up by the
tidal shock as
\beqar
\nonumber
N_{\rm T,s} & \approx & 7.5 \times 10^{66} 
\left( \frac{0.2 M_\odot}{M_{\rm Fe, SN}} \right) 
\left( \frac{y_{\rm Fe}}{y_{\rm Fe, \odot}} \right) 
\left( \frac{M_{\rm gas}}{10^9 M_\odot} \right)  \\
& \times &  \left( \frac{\eta_{\rm GCR}}{\eta_{\rm TCR}} \right)  
\left( \frac{n_{\rm ISM}}{1 {\rm cm^{-3}}} \right) 
\left( \frac{R_{\rm SNR}}{10 {\rm pc}} \right)^3 
\left( \frac{10^9 {\rm yr}}{\tau_{\rm TCR}} \right)
\eeqar

Finally, we estimate the amount of gas swept over by tidal shocks that
would yield the same level of lithium abundance as galactic
supernovae.
\beqar
M_{\rm T,s} & = & \mu N_{\rm T,s}   \\
\nonumber
& \approx &  8 \times 10^{9} M_\odot 
\left( \frac{0.2 M_\odot}{M_{\rm Fe, SN}} \right) 
\left( \frac{y_{\rm Fe}}{y_{\rm Fe, \odot}} \right) 
\left( \frac{M_{\rm gas}}{10^9 M_\odot} \right)  \\
&\times & \left( \frac{10^9{\rm yr}}{\tau_{\rm TCR}} \right) 
\left( \frac{n_{\rm ISM}}{1 {\rm cm^{-3}}} \right) 
\left( \frac{R_{\rm SNR}}{10 {\rm pc}} \right)^3 
\left( \frac{\eta_{\rm GCR}}{\eta_{\rm TCR}} \right) \\
\nonumber
\frac{M_{\rm T,s}}{M_{\rm gas}} &=& 8  
\left( \frac{0.2 M_\odot}{M_{\rm Fe, SN}} \right)
\left( \frac{y_{\rm Fe}}{y_{\rm Fe, \odot}} \right)    
\left( \frac{10^9{\rm yr}}{\tau_{\rm TCR}} \right)   \\
& \times & \left( \frac{n_{\rm ISM}}{1 {\rm cm^{-3}}} \right) 
\left( \frac{R_{\rm SNR}}{10 {\rm pc}} \right)^3 
\left( \frac{\eta_{\rm GCR}}{\eta_{\rm TCR}} \right)
\label{eq:main}
\eeqar
where we assumed the mean atomic mass $\mu = 1.3 m_{\rm H}$,
appropriate for the neutral ISM.

Equation~(\ref{eq:main}) indicates that in order for TCRs to
produce as much lithium as GCRs, up to a certain epoch in time characterized
by the solar metallicity, the entire galactic ISM must be tidally
shocked 8 times. For galactic encounters that can drive strong tidal
shocks in the interstellar medium of a "tidally harassed" satellite
galaxy, this would imply the occurrence of at least 8 close fly-bys.
However, even a single fly-by could result in a non-negligible
increase in lithium abundance in these galaxies. In the next
Section we describe the implications of this model for the Small
Magellanic Cloud.

\section{Implications for the Small Magellanic Cloud}

Adopting a SMC gas mass of $M_{\rm gas} (r <3 {\rm kpc})=3
\times 10^8 M_\odot$ \citep{bs}, the total energy required to pollute
all of the SMC gas with the solar level of \li6 abundance would be $E_6 \sim
10^{57}$ erg.  To estimate the kinetic energy of its galactic
encounters, we consider the interactions of the SMC with the Milky Way
and the Large Magellanic Cloud, given that both of these
have had significant gravitational impact on the SMC during its
history \citep[for e.g.,][]{diaz}. The total mass (including the dark
matter halo, gas, and stars) of the MW is $M_{\rm MW} \approx 10^{12}
M_\odot$ \citep{xue} and the total mass of the SMC is $M_{\rm SMC} (r<
3 {\rm kpc}) \approx 4 \times 10^9 M_\odot$ \citep{hz06}. The present day
separation of MW-SMC is $d=61$ kpc \citep{hildich} and, using 
equation (\ref{eq_kinetic}) we estimate the kinetic energy of their
encounter as $E_{kin} \approx 10^{58}$ erg. Thus, if the tidal interaction
of the SMC and the MW was to enrich the entire ISM of the SMC to a
solar metallicity value of \li6, less than 10\% of the kinetic energy
of the encounter at the current epoch would be used towards particle
acceleration. On the other hand, if we consider the LMC as the primary
tidal partner of the SMC, then with its total mass $M_{\rm LMC} (r< 9
{\rm kpc}) \approx 13 \times 10^9 M_\odot$, and 23 kpc present day
separation from the SMC \citep{besla,kva}, we estimate the total
kinetic energy from this interaction to be $E_{\rm kin} \approx 4
\times 10^{56}$ erg \footnote{Note that the equation~(\ref{eq_kinetic})
was derived for radially plunging orbits and it strictly does not
apply to a gravitationally bound pair of galaxies such as the SMC and
LMC -- in this context it only provides an estimate of the kinetic
energy within a factor of few.}. Consequently, the kinetic energy
between the LMC and SMC, as they are {\it today}, is insufficient to
account for a significant \li6 abundance production. The gravitational
interaction between the LMC and SMC has likely been much stronger
in the past, during their close approaches, and hence, could have
contributed to the total abundance of lithium in the SMC.  We discuss the
implications of the evolution of the SMC-LMC interaction over time in
Section~\ref{S_discussion}.

Since the observed metallicity of the SMC is approximately 1/5 solar
\citep{peimbert}, our model implies that tidal shocks would have to
sweep over the entire SMC ISM only about {\it twice} to accelerate
enough particles which would produce the same amount of lithium as the
GCRs. However, since any production of \li6 by GCRs must scale with
metallicity, it follows that $\li6_{\rm GCR}/\li6_\odot \approx
0.2$. For a typical GCR spectrum with spectral index $s=2.75$,
the production ratio between lithium isotopes from the same CR population
is $\li7 / \li6 \approx 1.3$ \citep{fp}. Thus, if TCRs have produced
the same amount of \li6 in the SMC as GCRs, this means that SMC
\li6 abundance should in fact be $\li6_{\rm SMC}/\li6_\odot \approx
0.4 $, while the isotopic ratio should be
\beqar
\nonumber
\left( \frac{\li7}{\li6} \right)_{\rm SMC} &=& 
\frac{\li7_{\rm BBN}+\li7_{\rm GCR}+\li7_{\rm TCR}+\li7_*}{\li6_{\rm GCR}+\li6_{\rm TCR}}\\
\nonumber
&=& \frac{\li7_{\rm BBN}+2 \times \li7_{\rm GCR}+\li7_*}{2 \li6_{\rm GCR}} \\
\nonumber
&=& \frac{\li7_{\rm BBN}+2 \times 1.3 \times \li6_{\rm GCR}+\li7_*}{2 \li6_{\rm GCR}} \\
& \approx & 10 + \epsilon_*
\label{equation15}
\eeqar
where $\epsilon_* \equiv \li7_*/(2\li6_{\rm GCR})$ is a small correction
to the lithium isotopic ratio that comes from the stellar production of
\li7.  For primordial and solar abundances we adopt $(\li7 /
H)_{\rm BBN} = 5.2 \times 10^{-10}$ \citep{cfo08} and $(\li6 /
H)_\odot = 1.53 \times 10^{-10}$ \citep{ag}, respectively.  Note that
the resulting ratio in equation~(\ref{equation15}) is almost a factor
of 2 smaller than the isotopic ratio $\sim18$ for the SMC, when GCRs
are considered to be the only post-BBN source of lithium. The value
obtained in equation~(\ref{equation15}) is consistent within error
with the best fit of the isotopic ratio recently obtained from
observations of the SMC by Hawk et al., who found an $(\li6 / \li7)_{\rm
SMC} = 0.13 \pm 0.05$ \citep{howk}. Note that our estimate of the
lithium isotopic ratio is not very sensitive to the precise nature of
the shocks and remains $(\li7 / \li6)_{\rm SMC} \approx 10$ even in
the case of cosmic rays with a spectral index $\alpha=2$ where lithium
isotopes are produced in a ratio $\li7 / \li6
\approx 2$. 

\section{Discussion}\label{S_discussion}

 In this section we discuss in more detail
the importance of the assumptions and parameters adopted in this
model. The key assumption of the model pertains to the unknown
spectrum of the tidal cosmic-ray population that arises in galactic
interactions. A physical property directly affected by this
uncertainty is $\epsilon_6$, the energy required to create one nucleus
of \li6, which in addition to the injection spectrum, also depends on
the cosmic-ray composition and confinement. For example, a lower
energy threshold would be obtained for systems where cosmic-ray
confinement is stronger (lower escape losses which results in a harder, less steep, propagated spectrum) 
and where metallicity is sufficiently high for
the production of \li6 through the CNO channel to become important.
It has been shown however, that for a wide range of plausible parameter
choices, $\epsilon_6$ has a value in the range of $5-100 \, \rm \, erg$
\citep{prantzos}. In case of the maximum energy threshold, the kinetic
energy of a close fly-by would be an order of magnitude below that
required to produce the solar lithium abundance. The implication is
that the fly-by model alone would fall short of explaining the
anomalously high abundance of \li6 in some galaxies, but could still
account for some non-negligible fraction of it.

Similarly, our estimate of the number of galaxy-galaxy encounters
capable of producing significant quantities of lithium also relies on the
assumption that the spectrum of the TCRs is indistinguishable from
the standard GCR spectrum in some galactic system.
Note however that a particular choice of cosmic-ray source composition and
form of cosmic-ray injection spectra (momentum vs. energy spectrum, see
\citep{prantzos} for discussion) apply to both cosmic-ray populations, and thus 
do not introduce additional degrees of freedom to our model.
The expression for spectrum and cosmic-ray composition, however, do affect the 
energy-per-nucleus threshold in the way discussed in the previous paragraph.

If the spectral indices of the two cosmic-ray populations are different (e.g.,
if the TCRs are accelerated in weaker shocks resulting in a softer, steeper,
spectra), this would result in a higher energy per nucleon requirement
for a steeper cosmic-ray spectrum, driven by the larger ionization losses
\citep{prantzos}. In that case, equation~(\ref{eq:main}) would
effectively depend on the ratio of fluxes $\Phi_{\rm GCR}/\Phi_{\rm
TCR}\propto E_{th}^{\alpha'}$ where $E_{th}$ is the threshold energy
and ${\alpha'}=\alpha_{\rm TCR} - \alpha_{\rm GCR}$.  
In our fiducial case we take this difference between spectra to be zero, and
thus the ratio of the fluxes comes down to the ratio between normalizations.
Related to this is our assumption of the instantaneous and constant TCR flux, where
we have omitted the unavoidable evolution of the TCR flux as this cosmic-ray
population is accelerated, and assumed that equilibrium flux is established. 
Given that TCRs are accelerated during the isolated events of close galactic fly-bys, it is probably
not true that TCR flux will reach equilibrium, thus evolution will have to be taken into
account. However, adopting this assumption is, for all practical purposes of this work,
equivalent to adopting a constant, mean TCR flux. 

The fundamental difference between tidal and SN shocks is their
physical scale -- tidal shocks in satellite galaxies operate on much
larger spatial scales than SNe. They can extend over a significant
fraction of the galaxy size, as traced by its stars and gas, and reach
scales over several kpc. We thus envision tidal shocks as large scale
SNR like structures. For a
single supernova, the cosmic-ray injection spectrum and maximal
acceleration energy depend on conditions like the blast
energy, local ISM density and properties of the magnetic field. On 
galactic scales however, a global GCR equilibrium spectrum is 
reached through the contribution of many supernova events throughout 
the galaxy. Thus, we only consider mean SNR properties, averaged 
over a number of SN events in comparison of TCR and GCR efficiency.
In other words, we assume that the two mechanisms 
operate under similar "global" conditions. Hence, as long as the velocity of the blast wave 
of the two processes is comparable over some stage of their evolution, 
their ability to accelerate charged particles should also be comparable.

An additional level of complexity may be
present due to the origin and evolution of the TCR population. While 
GCRs can reach an equilibrium between constant losses and
continuous injection over the lifetime of a system, the TCR spectrum
could reach an equilibrium only during epochs when large-scale
tidal shocks are actively propagating through the ISM and accelerating
particles. Once the particle injection ceases, TCRs would continue
to interact with the ISM, but their spectrum would be evolving rapidly
due to energy loses. In our work, we consider the equilibrium time
scale for TCRs comparable to the dynamical time of the interacting
system of galaxies. The time scale we adopt approximately accounts for strong encounters
, i.e., those capable of driving strong shocks and accelerating the TCRs, for which
the Mach number $\mathcal{M}> 100$. Our adopted value was estimated
for the specific encounters analyzed in this work and is essentially in agreement with
numerical simulations \citep{besla}; however this value can be in the range 
$\tau_{\rm TCR}\sim {\rm few}\times 10^8-10^9\,{\rm yr}$.
With respect to the limits of applicability, our
simple model, fails to explain any substantial level of \li6
abundance in galaxies when $\tau_{\rm TCR} \lesssim \tau_{\rm GCR}/100$.
In reality, the equilibrium time scale for the TCR flux depends on the
properties as well as the evolution of the ISM in a tidally
``harassed'' galaxy.  Clearly, careful numerical modeling of both
tidal shocks and particle acceleration is required for precise
determination of the resulting TCR spectrum; however, the purpose of
this work is to demonstrate the plausibility of this scenario, and we
defer the details to follow-up work. 

It is also worth noting that the SMC has experienced at least two close
encounters with the LMC, and is currently experiencing an ongoing
encounter with the MW \citep{diaz,besla}. The relative strength of
these interactions has varied as a function of time and orbital
parameters of the three galaxies. Cosmological models predict that
both the SMC and LMC could have been up to ten times more massive at
the time of their infall in the MW \citep{guo10,besla}. The Milky Way on
the other hand had a lower mass in the past than today, since its mass
increased over cosmic time. Simulations of cosmological
structure formation favor a scenario where the Magellanic Clouds are
currently on their first approach to the MW, thus implying that the
distance between the MCs and MW was larger in the past
\citep{besla}. All this points to a lesser role of the MW in tidal
interactions with the two satellites a few to ten billion years ago.
The same set of simulations finds that dwarf-dwarf galaxy interactions
of the SMC and LMC are the dominant driver of their evolution over the
past 5 -- 6~Gyr, during which they evolved as a gravitationally bound
pair.  During this time, the evolution of their baryonic component has
been dominated by tidal stripping and shocks. The SMC and LMC have
most likely had several close encounters with one another in the past,
during 2 -- 3 pericentric passages when their separation could have
been as small as a few~kpc. Given the larger masses and smaller
separation of the MCs in the past, it follows that the kinetic energy
of their interaction could have reached two orders of magnitude higher
values than that estimated for the SMC and MW system at the present
time.  If so, strong interactions of the SMC with LMC are likely to
have played a more important role for the acceleration of TCRs and
the production of lithium in both dwarf galaxies than their present day
interactions with the Milky Way. Given that over their cosmic history
the total mass of the LMC remained at least a few times larger than
that of the SMC, the LMC would have been less prone to tidal
harassment by its smaller companion and the Milky Way galaxy. Thus,
the past existence of the TCR population acting within Magellanic Clouds
can be tested by comparing lithium isotopic ratios in the Magellanic
Clouds. Specifically, a TCR population would have been more prominent
in the smaller interacting system, which implies a lower $\li7/\li6$
ratio in the SMC relative to the LMC. Different star-formation
histories of these two systems, on the other hand, resulted in an SMC
metallicity which is 0.2 of the solar, while the LMC metallicity is at the
level of 0.4 \citep{westerlund}. In the absence of TCRs from
both systems, from equation (\ref{equation15}) it follows that
the isotopic ratio would be lower in the LMC ($\li7/\li6
\approx 10$) compared to the SMC ($\li7/\li6
\approx 18$).  Therefore, if the lithium isotopic ratio was measured in
the LMC and was found to be comparable or higher than the SMC ratio
$(\li7/\li6)_{\rm SMC} \la (\li7/\li6)_{\rm LMC}$, this would be a
strong indication that a tidal cosmic-ray population was present (at
some epoch) within the SMC and has significantly impacted its chemical
evolution.

Although acceleration of the TCRs is independent of SNe, the presence of tidal shocks in 
the ISM may trigger star-formation, and result in an additional metallicity increase.
However, our model does not distinguish between secular star formation and that 
triggered by galaxy-galaxy interactions, but instead takes into account the integral number of 
supernova events over the history of the system. In this sense, our model is not sensitive to  
the exact star-formation history of the SMC. Therefore, even with
enhanced star-formation at some epoch, it can be shown that tidal
cosmic rays could have potentially produced lithium in quantities
comparable to what is expected from GCRs alone, resulting in
an anomalous lithium isotopic ratio.

\section{Conclusions}

Cosmic-ray nucleosynthesis is the dominant production channel of \li6
and one of the dominant sources of \li7, especially in higher
metallicity systems, where supernova remnants are taken as the
main acceleration sites of cosmic rays in star-forming galaxies.
In this work, we propose that tidal shocks which arise from close
galactic fly-bys can be an important source of cosmic rays, and thus
of lithium as well. Strong tidal shocks which could affect a significant
fraction of the gas content of a galaxy can occur in satellite systems
like the Small Magellanic Cloud, during its close fly-bys with the
Large Magellanic Cloud or the Milky Way. As a consequence, a
population of tidal cosmic rays that arises in satellite systems
can present an additional source of both lithium isotopes.  

The enrichment of SMC gas with extra lithium may bare important
consequences for the existing "lithium problem". Because of the discrepancy
between the predicted primordial lithium abundance and that measured in the
low-metallicity halo stars, it was suggested that lithium should be
measured in the gas phase of low metallicity systems. The first
measurement of this kind was recently carried out by \citet{howk}, in
SMC gas with a metallicity of $1/5$ solar, and is an important step
toward the resolution of this problem. The measured \li7 abundance,
is consistent with the expected primordial abundance, but is also just
marginally consistent with abundance expected for a system at 1/5 solar
metallicity where significant post-BBN lithium was produced in GCR interactions.
This marginal consistency means that there is
little room for post-BBN production of this isotope through stellar
process or cosmic-ray interactions. Therefore, with an additional
cosmic-ray population present, such as tidal cosmic rays, the
tension would be even greater, resulting in a discrepancy that is similar to
that observed in the low-metallicity halo star lithium.
Consequently such a scenario would indicate that the resolution
of the lithium problem is more likely to be found in non-standard
BBN. 

In case of the SMC, where Li has now been measured in the gas phase, our 
model shows that only two close fly-bys affecting the entire ISM of the SMC
are sufficient for the tidal cosmic rays to produce as much \li6 as
the galactic cosmic rays have produced over the entire history of the SMC.
Thus, given the already existing problem with the lithium abundances, and the
recent measurements of lithium in the gas phase of the SMC, which are consistent with
the predicted primordial abundance and above the observed lithium plateau values in halo stars,
it is crucial to test the fly-by hypothesis presented in this work, and to confirm that the SMC
is really a suitable environment for testing the lithium problem. On the other hand, if SMC gas
was enriched by additional lithium due to TCRs, this has to be taken into account and corrected for in order to
check the consistency of the SMC gas-phase lithium abundance with the expected primordial value.

As discussed in Section~\ref{S_discussion}, one possible test of the
presence of a new cosmic-ray population would be to compare the lithium
isotopic ratios between LMC and the SMC. Another possible approach is
based on the radio emissions of the interacting galaxies.  Tidal shocks
accelerate electrons to ultra-relativistic energies and provide
conditions for strong synchrotron radio-emission over relatively
short time-scales ($\sim 10^7$ yr). Therefore, an increase in radio
luminosity is expected in interacting systems, and especially in a
smaller member of the system. Indeed, a nearby interacting system
of galaxies, M51, shows an enhancement in radio-luminosity mostly
contributed by its smaller member, M51b, at low radio-frequencies,
that is two orders of magnitude higher relative to the unperturbed
galaxies. Such tidal interactions,
however, also lead to increased star-formation and GCR flux, and
consequently, enhanced radio-luminosity.  For this reason, suitable
candidates for testing the presence of TCRs with radio observations
would be galaxies in the early stage of interaction, or those that have
not reached the peak of fly-by driven star formation.
  
\begin{acknowledgments}
We are grateful to Brian D. Fields, Christopher J. Howk, Nicolas
Prantzos and Bojan Arbutina for their valuable comments and discussions.  We
are especially grateful to the anonymous Referees on their constructive comments
which helped make this paper better.  The work of T.P. is supported in
part by the Ministry of Education, Science and Technological
Development of the Republic of Serbia under project numbers 171002 and
176005.  Support for T.B. was in part provided by NASA through
Einstein Postdoctoral Fellowship Award Number PF9-00061 issued by the
Chandra X-ray Observatory Center, which is operated by the Smithsonian
Astrophysical Observatory for and on behalf of NASA under contract
NAS8-03060.  The work of D.U. is supported by the Ministry of
Education, Science and Technological Development of the Republic of
Serbia under project number 176005.
\end{acknowledgments}

{}

\end{document}